\documentclass[11pt,twoside]{article}

\usepackage{asp2006}
\usepackage{epsfig}

\begin{document}
\title{Cloud formation in giant planets}   
\author{Christiane Helling}   
\affil{SUPA, School of Physics \& Astronomy, Univ. of St Andrews, North Haugh, St Andrews,  KY16 9SS, UK}    

\begin{abstract} 
We calculate the formation of dust clouds in atmospheres of giant
gas-planets.  The chemical structure and the evolution of the grain
size distribution in the dust cloud layer is discussed based on a
consistent treatment of seed formation, growth/evaporation and
gravitational settling. Future developments are shortly addressed.
\end{abstract}


\section{Introduction}   
Clouds are a common feature in gas giant and brown dwarf atmospheres,
and observation start to spectroscopically infer their existence.  The
description of the actual formation of clouds in substellar
atmospheres has been a challenge in particular if consistently
calculated with radiative transfer and large-scale
hydrodynamics. Various branches pressed ahead in the recent years to
advanced the field of planet atmosphere modelling: i) detailed
radiative transfer calculations with simplified cloud chemistry
({\it Fortney} and {\it Barman} this volume), ii) large-scale multi-dimensional
hydrodynamical simulations but without dust chemistry ({\it Showman} and
{\it Menou}, this volume), iii) photochemical gas-phase models for a given
atmosphere stratification ({\it Tinetti}, this volume), and iv) detailed
non-equilibrium dust-cloud modelling for a prescribed atmosphere
structure. Our efforts are centred on point iv) which allows us to
study details of the cloud formation process like e.g. the cloud's
chemical composition and the grain size distribution function.  The
basic idea is that a phase-transition can only take place if the gas
is supersaturated. Supersaturation requires a considerable
super-cooling below the thermal stability temperature. Once seed
particles have formed (nucleation), chemical surface reactions quickly
form a mantle. The grain mantle consists of a variety of compounds in
an oxygen-rich environment because compounds can become thermally
stable in a very narrow temperature interval.  The newly formed dust
particles drift into deeper layers due to the high gravity of the
objects and grow further on their way into warmer regions. If the
temperature becomes too high, the compounds are not any more thermally
stable and, hence, they evaporate. Would these processes occur in a
truly static environment, no clouds would be present (Woitke \&
Helling 2004)\nocite{Woitke2004}. But substellar objects posses a
considerable convective over-shooting (Ludwig et al. 2006) which
brings up uncondensed material and which keeps the circle of dust
formation running.  Cloud formation due to the condensation of
solids/liquids reduces the remaining gas phase element abundances and
their gravitational settling introduces an additional dynamic process,
both changing the atmospheres appearance considerably.

In the following, we will demonstrate the results of our kinetic dust-cloud
model for gas-giant atmospheres and shortly address some future developments.

\begin{figure}[h]
   \centering
  \epsfig{file=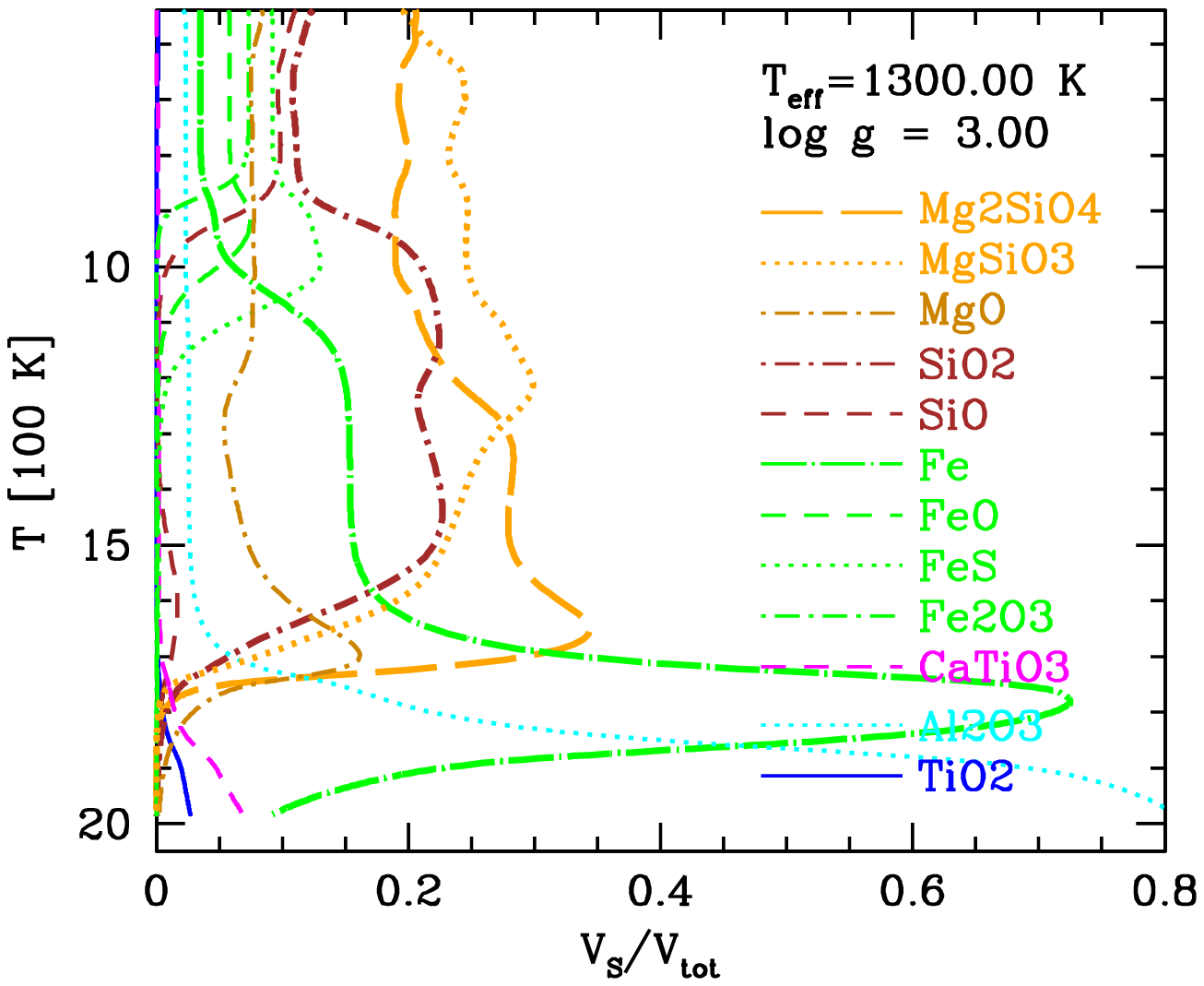, scale=0.9}\\[0.2cm]
  \epsfig{file=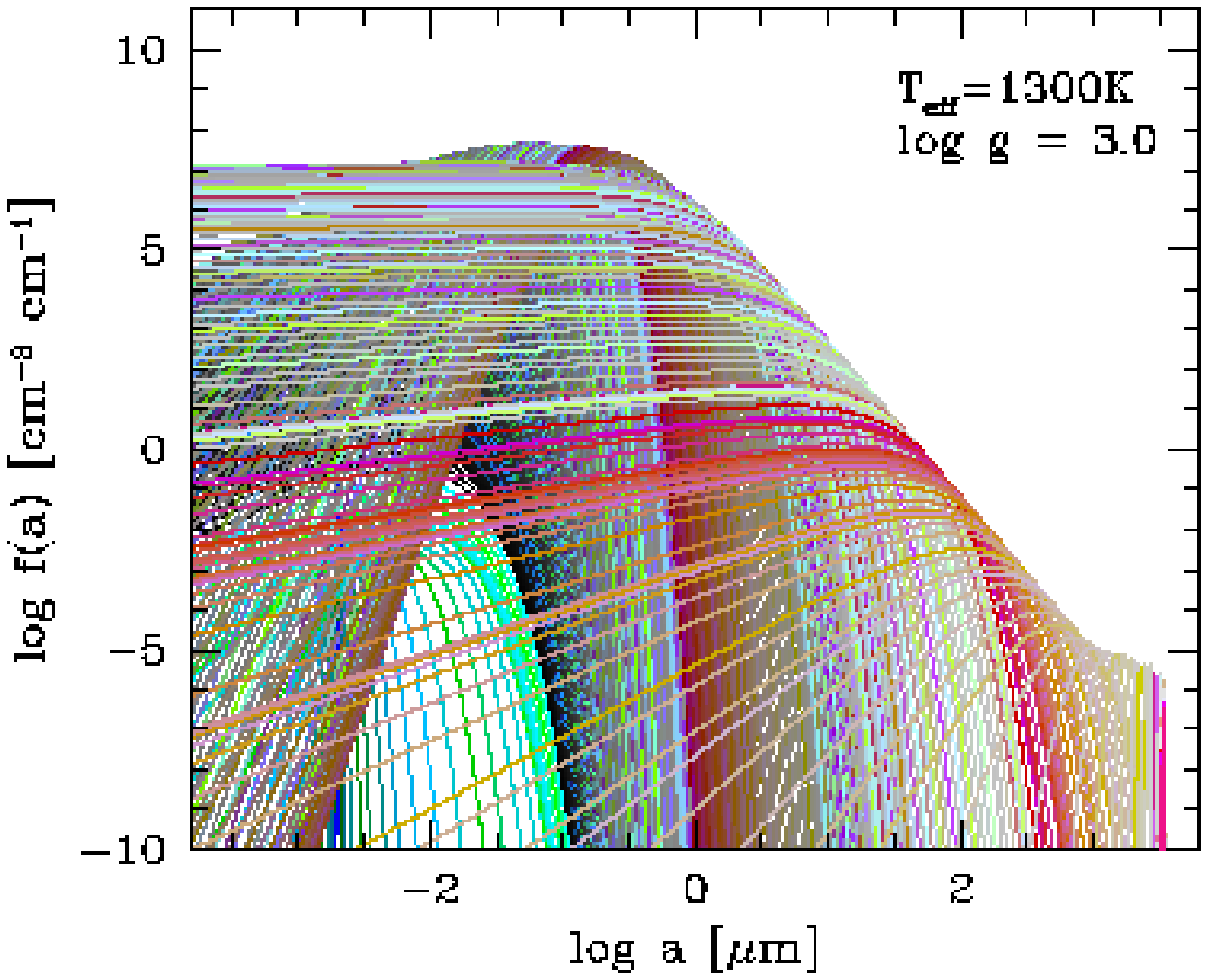, scale=0.9}
   \vspace*{-0.1cm}
      \caption{Atmospheric dust cloud in a giant gas-planet with  T$_{\rm eff}=1300\,$K, $\log$\,g$=3.0$.
{\bf Top:} Material composition (in volume fractions) of the dust cloud.
 {\bf Bottom:} Grain size distribution functions for each atmospheric layer in the dust cloud.
              }
  \label{meanSize}
   \vspace*{-0.3cm}
   \end{figure}

\section{Method}
We model nucleation (seed formation), heterogeneous growth,
evaporation, and drift (gravitational settling) of dirty dust
particles in a quasi-static atmosphere by using the moment method
(Gail\& Sedlmayr 1988, Dominik et al 1993, Woitke\& Helling 2004,
Helling \& Woitke 2006; Helling, Woitke \& Thi
2007)\nocite{Helling2005}\nocite{Helling2007}. We consider the
formation of compact spherical grains in an oxygen-rich gas by the
initial nucleation of TiO$_2$ seed particles, followed by the growth
of a dirty mantle.  The moment and elemental conservation equations
are evaluated for given $(T,\rho, v_{\rm conv})$ either for a
prescribed static model atmosphere structure or inside a iterative
solution of the radiative transfer problem.  Our dust model calculates
the amount of condensates, the mean grain size $\langle a \rangle$,
the parameterised grain size distribution function, and the volume
fractions $V_{\rm s}$ of each material as a function of height $z$ in
the atmosphere. 12 solids made of 8 elements are considered to form
the grain mantle by 60 chemical surface reactions. We solve 19 stiff
differential equations already in the dust and element conservation
complex.

\section{Results}

\subsection{Dust composition in atmospheres of gas-giant  planets}

Figure~\ref{meanSize} (top) shows the cloud's material composition for
a gas giant.  The upper cloud regions are populated by small silicate
grains which are composed mainly of Mg$_2$SiO$_4$/MgSiO$_3$ and
SiO/SiO$_2$. Iron-binding solids are thermally stable and contribute
in total as much as the Mg-binding compounds do. Two instability
regions occur where the cloud's material composition changes
considerably: One is where the iron compounds and SiO evaporate, and a
second one occurs at higher temperatures where all the remaining
silicates evaporate. We note a stratified purification of the dust
cloud. The warmest dust cloud layers are mainly made of Fe[s] and
Al$_2$O$_3$[s] with little inclusions of TiO$_2$[s] and
CaTiO$_3$[s]. Grains forming the cloud base have large mean grain
sizes and the grain size distribution $f(a)$ is very narrow.
Figure~\ref{meanSize} (bottom) shows that $f(a)$ is
$\delta$-function-like at the very right of the plot (i.e. large grain
size $a$) relating to the cloud base.

\subsection{Grain size distributions in atmospheres of gas-giant  planets}

The grain size distribution $f(a)$ is based on the solution of our
moment equations and is here parameterised as a potential exponential
size distribution.  Figure ~\ref{meanSize} (bottom) demonstrates
 that the size distribution changes with height in the
atmosphere. The l.h.s. of the plot (blueish) shows the nucleation
regime and $f(a)$ is $\delta$-function-like also at the cloud deck. It
broadens and increases if nucleation and growth run in parallel. If
nucleation ceases, $f(a)$ moves through the grain size space towards
larger grains until evaporation sets in. Evaporation causes the
smallest grains to disappear, hence, $f(a)$ decreases and
narrows. Evaporation also decreases the size of existing grains,
hence, $f(a)$ broadens considerably towards smaller grain sizes.  We
have, hence, shown that it is difficult to attribute a single grain
size to a cloud and that the shape of the grain size distribution
varies through the entire cloud.

\section{The future}
Detailed micro-physical models of cloud formation allow the study of
the actual formation and evolution of the atmospheric cloud
constituents. It is, however, deemed challenging to couple a complete
micro-physical model consistently with atmosphere simulations and
still preserve its flexibility.  Additionally, the coupling of dust
{\it formation} and multi-dimensional hydro-simulations needs to deal
with the time-scale problem between chemistry and fluid dynamics, and
with the turbulent closure problem. However, Dehn et al. (2007) did
combine a reduced version of our kinetic cloud formation model with
the 1D {\sc Phoenix} model atmosphere code by Hauschildt \& Baron
(1999) and first consistent simulations become available.

Generally, the solution of the stellar/planetary atmosphere problem
should be determined by only the stellar/planetary parameter and
ideally, no parameter would be attributed to clouds, day-nigh-effects
etc. The hope is that the solution of the atmosphere problem
(classically: hydrostatic equilibrium, radiative transfer, mixing
length theory, gas-phase chemical equilibrium + cloud model) is
unique, and hence, the resulting synthetic spectrum is determined by
this stellar/planetary parameter combination, too. It is, however,
noticeable difficult to identify a single stellar/planetary parameter
combination that fits an observed spectral range (e.g. Brandecker et
al. 2006). To advance this situation, modellers have set out to
conduct a comparative study of cloud models in
order to provide a measure for uncertainties inherent to substellar
model
atmospheres\footnote{http://www.lorentzcenter.nl/lc/web/2006/203/info.php3?wsid=203}.


\end{document}